\documentclass[a4paper, 12pt]{article}
\usepackage[T1]{fontenc}
\usepackage[utf8]{inputenc}
\usepackage{amsfonts}
\usepackage{amssymb}
\usepackage{indentfirst}
\usepackage{amsmath}
\usepackage{amsthm}
\usepackage[pdftex]{graphicx}

\begin{document}
\bibliographystyle{plain}
\begin{center}
Marta Dudek\\
Institute of Mathematics, University of Szczecin (Szczecin, Poland)\\
(marta.dudek@vp.pl)
\end{center}
\begin{center}
Janusz Garecki\\
Institute of Mathematics and Cosmology Group, University of Szczecin (Szczecin, Poland)\\
(janusz.garecki@usz.edu.pl)
\end{center}
\begin{center}
\begin{large}
\textbf{RIEMANNIAN STRUCTURE IMPOSED ON FRIEDMANN AND MORE GENERAL SPACETIMES}
\end{large}
\end{center}
\hspace{5cm}In memory of Professor Julian Ławrynowicz
\\
\\
\textbf{Summary}

In the paper we consider two Finsler-like Riemannian metrics, which can be in a natural way introduced into general relativity. One of those metrics $\gamma _{ab}$ is degenerate and the second $h_{ab}$ is nondegenerate. We are mainly interested with the metric $h_{ab}$ and comparing the geometric structure determined by this metric $h_{ab}$ with the geometric structure determined by the Lorentzian metric $g_{ab}$ of the underlying spacetime. Full comparison we have given for Friedmann Universis. 

The preliminary version of the paper was presented by one of us (J.G.) on conference POTOR 6 in Szczecin 2019. We think that the our introduction of the Riemannian metric $h_{ab}$ into spacetime is simpler and more general than to so-called Wick rotation.

\newpage
\section*{Introduction}
Nowadays the mathematical model of the physical spacetime is given by real 4-dimentional Lorentzian (or pseudoriemannian) manifold which metric satisfies Einstein equations.

In quantum field theory (QFT) the Lorentzian structure leads to serious complications in calculation of the so-called "sums over trajektories". These sums are improper integrals over spacetime. The calculation of these integrals is essentially simplified in the framework of the Riemannian geometry. So, specialists on QFT apply trick called "Wick rotation". This trick relies on introduction complex time $u=t+i\tau $ and on rotation in complex plane $U$ from real axis $t$ to imaginary axis $i\tau $.

Such Wick rotation changes Lorentzian metric onto Riemannian metric and Lorentzian manifold into Riemannian. After performing calculations in so obtained Riemannian geometry they finally come back to prevoius Lorentzian geometry by using analytical extension.

Unfortunately, Wick rotation has clear physical meaning only in the case of flat Lorentzian spacetime, i.e., in Minkowskian spacetime. In the case of the general relativistic spacetimes such procedure is physically unclear because we have no distinguised axis of time in this case.

In our approach we propose another way how to change Lorentzian geometry onto Riemannian. We have called obtained Riemannian geometry "Riemannian structure imposed on Lorentzian".

Our approach uses only real quantities and is more general then the Wick rotation. It can be applied to any solution of the Einstein equations.

\newpage
\section{Riemannian, Finsler-like, metrics used in GR.}
One can consider at least the two Riemannian, Finsler-like metrics in GR:
\begin{align}
\gamma_{ab}:=v_{a}v_{b}-g_{ab}=\gamma_{ab}(x;\vec{v}),
\\h_{ab}=h_{ab}(x;\vec{v}):=2v_{a}v_{b}-g_{ab}(x).
\end{align}
Here $g_{ab}=g_{ab}(x)$ is the Lorentzian metric of a spacetime and $\vec{v}$ is an unit timelike vector field: $\big \| \vec{v}\big \|=\sqrt{g_{ab}v^{a}v^{b}}=1$.\\
Physically $\vec{v}$ is the four-velocity of an observer $O$.\\
\underline{Remark:} An observer in GR is defined as the unit timelike vector $\vec{v}$ tangent to the timelike world-line $x^{a}=x^{a}(s)$, where $s=$ natural parameter along this world-line.\\
After fixing vector field $\vec{v}$, i.e., after fixing observer $O$, one has in the adapted coordinates $(x^{b})$
\begin{align}
v^{a}\stackrel{*}{=}\frac{\delta ^{a}_{0}}{\sqrt{g_{00}}}\implies v_{a}\stackrel{*}{=}\frac{g_{a0}}{\sqrt{g_{00}}} \notag \\
\gamma_{ab}\stackrel{*}{=}\Bigl (\frac{g_{a0}g_{b0}}{g_{00}}-g_{ab} \Bigr )=\gamma _{\alpha \beta} \\
h_{ab}\stackrel{*}{=}\Bigl (\frac{2g_{a0}g_{b0}}{g_{00}}-g_{ab} \Bigr ).
\end{align}
(3)-(4) are Riemannian metrics.
$\stackrel{*}{=}$ means that equality is valid in special coordinates only (adopted coordinates).\\
$\gamma_{ab}(x)=\gamma_{ba}(x)$ (with a fixed $\vec{v}$) is a degenerate Riemannian metric: $\gamma_{ab}=diag(0,1,1,1)$.
From the physical point of view the $\gamma_{ab}(x) $ (with a fixed $\vec{v}$) is a 3-dimensional spatial metric in a (near) vicinity of the observer $O$ (which velocity is $\vec{v}$) which determines distances by using radar method, i.e., by sending and detecting reflected light signals.\\
Namely, one can easily show [1] that the spatial line element for the observer $O$ reads
\begin{align}
dl^{2}=\gamma_{ab}dx^{a}dx^{b}\stackrel{*}{=}\gamma_{\alpha \beta}dx^{\alpha }dx^{\beta }\stackrel{*}{=}-g_{\alpha \beta }dx^{\alpha }dx^{\beta } \\
v^{a}\stackrel{*}{=}\frac{\delta ^{a}_{0}}{\sqrt{g_{00}}}\implies v_{a}\stackrel{*}{=}\frac{g_{a0}}{\sqrt{g_{00}}},\notag \\
\alpha ,\ \beta ,\ \gamma , ...=1,\ 2,\ 3) \notag \\
a,\ b,\ c,... =0,\ 1,\ 2,\ 3 ) \notag
\end{align}
So, $\gamma _{ab}$, determines distances in a vicinity of $O$ as measured by radar method.On the other hand, $h_{ab}\stackrel{*}{=}diag(1,1,1,1)$ is a 4-dimentional positive-definite Riemannian metric (after fixing $\vec{v}:\ v^{i}v_{i}=1$). It determines Riemannian structure which is tangent to Finsler-like structure determined by $h_{ab}(x;\vec{v})$ with variable $\vec{v}$.\\
A possible physical meaning of the Finsler-like metric $h_{ab}=2v_{a}v_{b}-g_{ab}$ and its applications are still under investigation. This metric revealed at the first time in investigations on superenergy and supermomentum tensors in GR [2].\\
\underline{Remark} Observe that the both metrics $\gamma_{ab}=v_{a}v_{b}-g_{ab}$ and $h_{ab}=2v_{a}v_{b}-g_{ab}=v_{a}v_{b}+\gamma _{ab}$ are uniquely determined along the world-line $x^{a}=x^{a}(s)$ of a fiducial observer $O$(and then they are Riemannian metrics).\\
\underline{In the following we confine to the Finsler-like metric $h_{ab}=h_{ba}(x;\vec{v})$.}

\newpage
\section{The general properties of the Riemannian Finsler-like metric $h_{ab}=2v_{a}v_{b}-g_{ab}$}

1.\ \ For the fixed observer $O$ $h_{ab}(x)=h_{ba}(x)$ is an uniquely determined and positive-definite Riemannian metric with signature $(++++)$
\begin{align}
h_{ab}(x)=2v_{a}v_{b}-g_{ab}\stackrel{*}{=}\Bigl (\frac{2g_{a0}g_{b0}}{g_{00}}-g_{ab}\Bigr )\\ \notag
v^{a}\stackrel{*}{=}\frac{\delta ^{a}_{0}}{\sqrt{g_{00}}} \\ \notag
v_{a}\stackrel{*}{=}\frac{g_{a0}}{\sqrt{g_{00}}}\\ 
\end{align}

$h_{ab}(x)$ determines Riemannian structure tangent to the Finsler-like structure determined by $h_{ab}(x;\vec{v})$.\\
2.\ \ $h_{ab}(x)$ preserves norm of the four-velocity $\vec{v}$ of the observer $O$: $h_{ab}v^{a}v^{b}=1=||\vec{v}||^{2}$.\\
3. \ \ Causal structure of the spacetime can be expressed in term of $h_{ab}(x)$, e.g., \\
$\widetilde{ds}^{2}=h_{ab}dx^{a}dx^{b}<2v_{a}v_{b}dx^{a}dx^{b}$ (timelike intervals $ds^{2}>0$\\
$\widetilde{ds}^{2}=h_{ab}dx^{a}dx^{b}=2v_{a}v_{b}dx^{a}dx^{b}$ (null intervals $ds^{2}=0$\\
$\widetilde{ds}^{2}=h_{ab}dx^{a}dx^{b}>2v_{a}v_{b}dx^{a}dx^{b}$ (spatial intervals $ds^{2}<0$\\
where $\widetilde{ds}^{2}>0$ and
\begin{align}
ds^{2}=g_{ab}dx^{a}dx^{b} 
\end{align}
is an interval in a Lorentzian spacetime $(M_{4};g)$ of GR [$g_{ab}$ possesses signature $(+---)$].

In the Friedman spacetimes (like as in a static spacetime) we have a physically distinguished unit timelike vector field $\vec{v}:\ v^{i}v_{i}=1$ - the four-velocity field of the so-called "isotropic observers". So, in this case, one can introduce univocally the Riemannian metric $h_{ab}=2v_{a}v_{b}-g_{ab}$ globally, on the whole spacetime.
We do this and in the following we will compare the ordinary Lorentzian structure of the Friedman spacetimes with the geometric structure imposed on these spacetimes by the Riemannian metrics $h_{ab}$ and $h$.\\
The comparison will be done in the coordinates $x^{0}=t$, $x^{1}=\chi $, $x^{2}=\vartheta $, $x^{3}=\varphi [1]$.

\underline{\textbf{Lorentzian structure determined by $g_{ab}$}} \\
\underline{Line element}
\begin{align}
&ds^{2}=g_{ab}dx^{a}dx^{b}\stackrel{*}{=}dt^{2}-R^{2}(t)[d\chi ^{2}+S^{2}(\chi )(d\vartheta ^{2}+sin^{2}\vartheta d\varphi ^{2})]
=dt^{2}-dl^{2},
\end{align}
where 
$$
S(\chi )=\left\{ \begin{array}{ll}
sin \chi & \textrm{if $k=1$}\\
\chi &\textrm{if $k=0$}\\
sh \chi &\textrm{if $k=-1.$}
\end{array}\right.
$$
$k$ is the so-called normalized curvature of the spatial slices $t=const$.\\
\underline{Spatial line element}
\begin{align}
dl^{2}=\gamma _{\alpha \beta }dx^{\alpha }dx^{\beta }=R^{2}(t)[d\chi ^{2}+S^{2}(\chi )(d\vartheta ^{2}+sin^{2}\vartheta d\varphi ^{2})]
\end{align}
\underline{Intervals}\\
timelike
\begin{align}
ds^{2}>0 \notag
\end{align}
null
\begin{align}
ds^{2}=0 \notag
\end{align}
spacelike
\begin{align}
ds^{2}<0
\end{align}
\underline{Raising and lowering of indices}
\begin{align}
&u_{a}=g_{ab}u^{b}, \notag
\\&u^{a}=g^{ab}u_{b}
\end{align}
\underline{Proper time}
\begin{align}
d\tau ^{2}=g_{00}dx^{0^{2}}\implies d\tau =\sqrt{g_{00}}dx^{0}\stackrel{*}{=}dt
\end{align}
\underline{Levi-Civita's connection (intrinsic, non-zero components)}
\begin{align}
&\Gamma ^{0}_{11}=R\dot{R},\ \ \Gamma^{2}_{12}=\Gamma ^{3}_{13}=\frac{S'}{S}, \notag
\\&\Gamma ^{1}_{10}=\Gamma ^{2}_{20}=\Gamma ^{3}_{30}=\frac{\dot{R}}{R}, \ \ \Gamma^{1}_{22}=-SS', \notag
\\& \Gamma ^{0}_{22}=R\dot{R}S^{2},\ \ \Gamma ^{3}_{23}=ctg\vartheta ,\ \ \Gamma ^{1}_{33}=\Gamma ^{1}_{22}sin^{2}\vartheta , \notag
\\&\Gamma ^{2}_{33}=-\frac{sin2\vartheta }{2},\ \ \Gamma ^{0}_{33}=\Gamma ^{0}_{22}sin^{2}\vartheta , \notag
\\&\dot{R}:=\frac{dR}{dt}, \ \ S':=\frac{dS}{d\chi }.
\end{align}
\underline{Riemann's tensor components (intrinsic, non-zero components)}
\begin{align}
&R_{0101}=R\ddot{R}, \ \ R_{0202}=R_{0101}S^{2},\notag
\\&R_{0303}=R_{0202}sin^{2}\vartheta , \ \ R_{1212}=R^{2}S(S''-S\dot {R}^{2}), \notag
\\&R_{1313}=R_{1212}sin^{2}\vartheta , \ \ R_{2323}=R^{2}S^{2}sin^{2}\vartheta (S^{2}-\dot{R}^{2}S^{2}-1)
\end{align}
\underline{Symmetries (Killing fields)}\\
There exist 6 spatial Killing vector fields $\xi ^{i}_{A}$, $A=1,...,6$ which satisfy Killing equations
\begin{align}
\nabla _{(i}\xi _{k)}=0
\end{align}\underline{Einstein tensor (non-zero components)}
\begin{align}
&G^{00}=\frac{3\dot{R}^{2}}{R^{2}}-\frac{2S''}{R^{2}S}-\frac{s'^{2}}{R^{2}S^{2}}+\frac{1}{R^{2}S^{2}}\notag
\\&G^{11}=\frac{1}{R^{2}}\Bigl (\frac{-1}{R^{2}S^{2}}-\frac{\dot{R}^{2}}{R^{2}}-\frac{2\ddot{R}}{R}+\frac{S'^{2}}{R^{2}S^{2}}\Bigr ) \notag
\\&G^{22}=\frac{1}{R^{2}S^{2}}\Bigl (\frac{S''}{R^{2}S}-\frac{2\ddot{R}}{R}-\frac{\dot{R}^{2}}{R^{2}}\Bigr ) \notag
\\&G^{33}=\frac{1}{sin^{2}\theta }G^{22}
\end{align}
\underline{Geodesic lines}\\
(We cite only the one equation which differs for the considered two metrics. The remaining three equations are identic.)
\begin{align}
&\ddot{t}+RR'\Bigl [\frac{\dot{r}^{2}}{1-kr^{2}}+r^{2}\dot{\vartheta }^{2}+r^{2}sin^{2}\vartheta  \dot{\varphi }^{2}\Bigr ]=0
\\& \dot{t}:=\frac{dt}{du}\notag
\\&\dot{r}:=\frac{dr}{du}\notag
\\&\dot{\vartheta }:=\frac{d\vartheta }{du}\notag
\\&\dot{\varphi }:=\frac{d\varphi }{du}\notag
\\&R':=\frac{dR}{dt}\notag
\\&\dot{R}=\frac{dR}{dt} \notag
\end{align}
where $u=$ an affine parameter, $r=S(\chi )$.\\
$r=S(\chi )$ [Here we have used the comoving coordinates ($t,r,\vartheta ,\varphi $)] in which $ds^{2}=dt^{2}-R^{2}(t)\Bigl [\frac{dr^{2}}{1-kr^{2}}+r^{2}(d\vartheta ^{2}+r^{2}sin^{2}\vartheta d\varphi ^{2})\Bigr ]$.\\

\underline{\textbf{Structure determined by Riemannian metric $h_{ab(x;\vec{v})}$ with fixed $\vec{v}$}}\\
\underline{Line element}\\
\begin{align}
 \widetilde{ds}^{2}=h_{ab}dx^{a}dx^{b}\stackrel{*}{=}dt^{2}+R^{2}(t)[d\chi ^{2}+S^{2}(\chi)(d\vartheta ^{2}+sin ^{2}\vartheta d\varphi ^{2})]=dt^{2}+dl^{2}
\end{align}
\underline{Spatial line element}
\begin{align}
dl^{2}=\gamma _{\alpha \beta }dx^{\alpha }dx^{\beta }\stackrel{*}{=}R^{2}(t)[d\chi ^{2}+S^{2}(\chi)(d\vartheta ^{2}+sin ^{2}\vartheta d\varphi ^{2})]
\end{align}
\underline{Intervals:}\\
timelike
\begin{align}
\widetilde{ds}^{2}<2v_{a}v_{b}dx^{a}dx^{b}\stackrel{*}{=}2dx^{0^{2}}=2dt^{2} \notag
\end{align}
null
\begin{align}
\widetilde{ds}^{2}=2v_{a}v_{b}dx^{a}dx^{b}\stackrel{*}{=}2dt^{2} \notag
\end{align}
spacelike
\begin{align}
\widetilde{ds}^{2}>2v_{a}v_{b}dx^{a}dx^{b}\stackrel{*}{=}2dt^{2}
\end{align}
\underline{Raising and lowering of indices}
\begin{align}
&\bar{u}_{a}=h_{ab}u^{b}\rightarrow \bar{u}_{0}\stackrel{*}{=}u_{0}, \ \ \bar{u}_{\alpha }\stackrel{*}{=}-u_{\alpha }\notag
\\&\bar{u}^{a}=h^{ab}u_{b}\rightarrow \bar{u}^{0}\stackrel{*}{=}u_{0},\ \ \bar{u}^{\alpha }\stackrel{*}{=}-u^{\alpha }
\end{align}
\underline{Proper time}\\
\begin{align}
&d\bar{\tau}^{2}=h_{00}dx^{0^{2}}\stackrel{*}{=}g_{00}dx^{0^{2}}\notag
\\&d\bar{\tau }\stackrel{*}{=}\sqrt{g_{00}}dx^{0}=dt \notag
\end{align}
\underline{Levi-Civita's connection (intrinsic, non-zero components)}
\begin{align}
&\Gamma ^{0}_{11}=-R\dot{R},\ \ \Gamma^{2}_{12}=\Gamma ^{3}_{13}=\frac{S'}{S}, \notag
\\&\Gamma ^{1}_{10}=\Gamma ^{2}_{20}=\Gamma ^{3}_{30}=\frac{\dot{R}}{R}, \ \ \Gamma^{1}_{22}=-SS', \notag
\\& \Gamma ^{0}_{22}=-R\dot{R}S^{2},\ \ \Gamma ^{3}_{23}=ctg\vartheta ,\ \ \Gamma ^{1}_{33}=\Gamma ^{1}_{22}sin^{2}\vartheta , \notag
\\&\Gamma ^{2}_{33}=-\frac{sin2\vartheta }{2},\ \ \Gamma ^{0}_{33}=\Gamma ^{0}_{22}sin^{2}\vartheta , \notag
\\&\dot{R}:=\frac{dR}{dt}, \ \ S':=\frac{dS}{d\chi }.
\end{align}
\underline{Riemann's tensor components (intrinsic, non-zero components)}
\begin{align}
&R_{0101}=-R\ddot{R}, \ \ R_{0202}=R_{0101}S^{2},\notag
\\&R_{0303}=R_{0202}sin^{2}\vartheta , \ \ R_{1212}=-R^{2}S(S''-S\dot{R}^{2}), \notag
\\&R_{1313}=R_{1212}sin^{2}\vartheta , \ \ R_{2323}=-R^{2}S^{2}sin^{2}\vartheta (S'^{2}-\dot{R}^{2}S^{2}-1)
\end{align}
\underline{Symmetries (Killing fields)}\\
Riemannian metric $h_{ab}$ has the same symmetries as Lorentzian metric $g_{ab}$ i.e., it admints the same  6 Killing fields $\xi ^{i}_{A}$, $A=1,...,6$ admitted by $g_{ab}$.\\
\underline{Einstein tensor (non-zero components)}
\begin{align}
&G^{00}=\frac{2S''}{R^{2}S}-\frac{S'^{2}}{R^{2}S^{2}}+\frac{3\dot{R}^{2}}{R^{2}}+\frac{1}{R^{2}S^{2}}, \notag
\\&G^{11}=\frac{1}{R^{2}}\Bigl (\frac{1}{R^{2}S^{2}}-\frac{\dot{R}^{2}}{R^{2}}-\frac{2\ddot{R}}{R}+\frac{S'^{2}}{R^{2}S^{2}}\Bigr ) \notag
\\&G^{22}=\frac{1}{R^{2}S^{2}}\Bigl (-\frac{S''}{R^{2}S}-\frac{2\ddot{R}}{R}-\frac{\dot{R}^{2}}{R^{2}}\Bigr ) \notag
\\&G^{33}=\frac{1}{sin^{2}\theta }G^{22}
\end{align}
\underline{Geodesic lines}\\
\begin{align}
\ddot{t}-RR'\Bigl [\frac{\dot{r}^{2}}{1-kr^{2}}+r^{2}\dot{\vartheta }^{2}+r^{2}sin^{2}\vartheta \dot{ \varphi }^{2}\Bigr ]=0
\end{align}
The remaining three equations are identic with those for $g_{ab}$.

The isotropic observers for which $v^{i}\stackrel{*}{=}\delta ^{i}_{0}$, $u=s=t$, $r=$const, $\vartheta =$const, $\varphi =$const, satisfy trivially the geodesic eqations in both structures.\\
Of course, the dynamical evolution of the universes, i.e., slices $t=$const, predicted by Einstein equations is the same in both metrics $g_{ab}$ and $h_{ab}=2v_{a}v_{b}-g_{ab}$ (with fixed $\vec{v}:\ v^{a}\stackrel{*}{=}\delta ^{a}_{0}$).

\underline{Results of the above comparison:}
\begin{enumerate}
\item Spatial line elements are the same.
\item Proper time is identic in both metrics.
\item The absolute values of the Levi-Civita's connection are the same in both metrics. The same is true for Riemann tensors components.
\item Both metrics admit the same Killing vector fields.
\item The components of the Einstein's tensors are only slightly different.
\item For geodesic lines we have only one difference in sign of the second terms of the cited equations. The remaining three equations are identic.
\end{enumerate}

\newpage
\section{Remarks on the more general solutions to the Einstein equations}

In general, one can express any solution to the Einstein equations in normal Gauss coordinates (NGC) $(s, \xi ^{\alpha })$ (See Fig.1).\\

\begin{figure}[bh]
\centering
\includegraphics[width=1.0\textwidth]{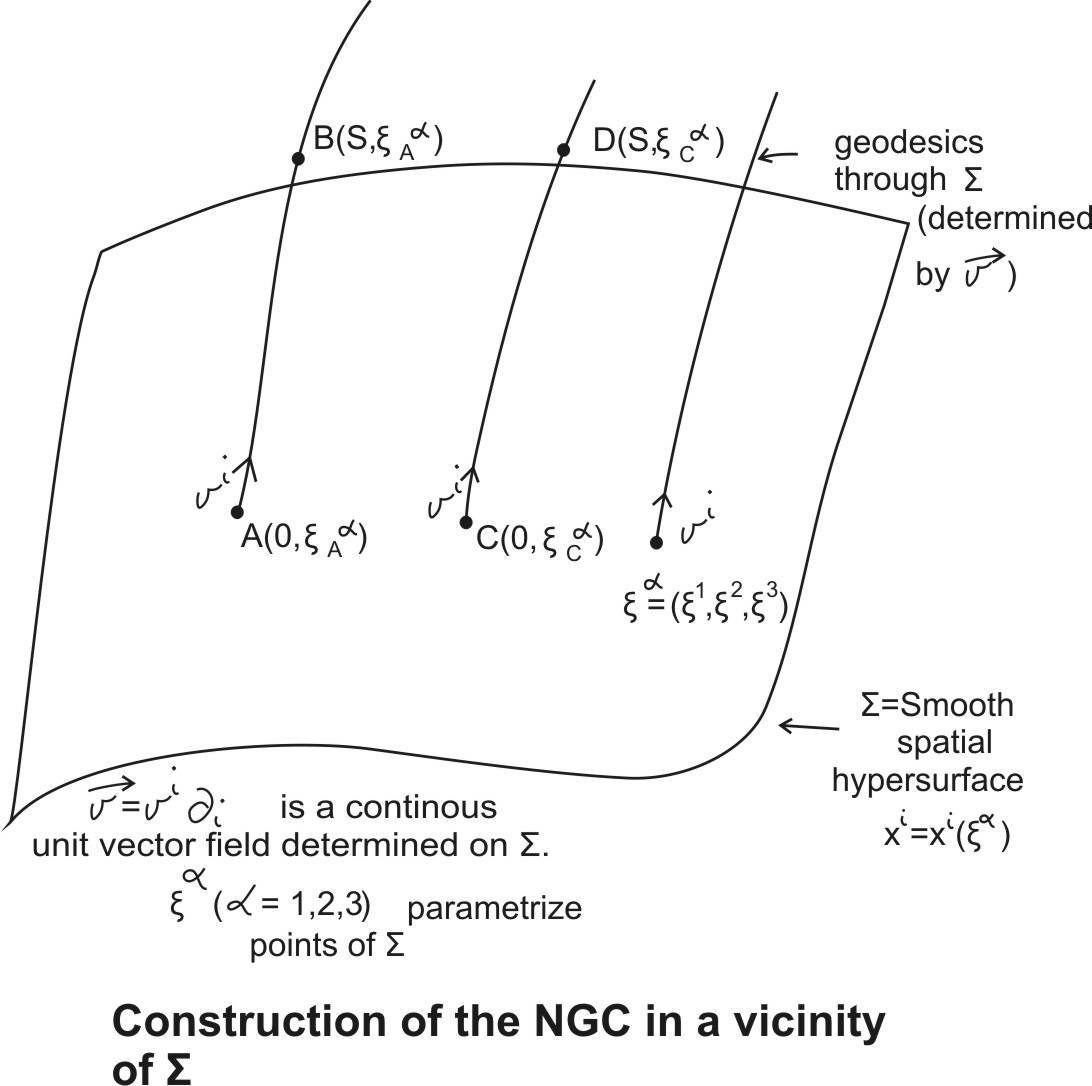}
\caption{Construction of the NGC in a vicinity of $\Sigma $}
\end{figure}

NGC of $B:=(s;\xi ^{\alpha }_{A})$, where $s$ is a natural affine parameter along geodesic which connects $A\in \Sigma $ and $B$ ($s=0$ on $\Sigma $), $k^{\alpha }\ (\alpha =1,2,3)$ parametrize points of $\Sigma$.\\
In NGC we have the situation very like to the considered earlier in Friedman case. [It is because the coordinates $(t, \chi , \vartheta , \varphi )$ or $(t, r, \vartheta , \varphi )$ used in there were also NGC (and simultaneously comoving).] Namely, we have a distinguished and unit timelike vector field $\vec{v}$ tangent to the geodesics through initial and spatial $\Sigma $.\\
Moreover, in the NGC we have [like as in the comoving NGC $(t, \chi , \vartheta , \varphi )$ in Friedman case]
\begin{align}
g_{00}=g^{00}=1,\ \ g_{0\alpha }=g^{0\alpha }=0 
\end{align}
and 
\begin{align}
ds^{2}=g_{ab}dx^{a}dx^{b}\stackrel{*}{=}dt^{2}+g_{\alpha \beta }dx^{\alpha }dx^{\beta }=dt^{2}-\gamma _{\alpha \beta }dx^{\alpha }dx^{\beta }, 
\end{align}
where 
\begin{align}
\gamma _{ab}=v_{a}v_{b}-g_{ab}\stackrel{*}{=}g_{a0}g_{b0}-g_{ab}=-g_{\alpha \beta }
\end{align}
is a Finsler-like spatial metric (with fixed $\vec{v}$).

$\vec{v}=v^{a}\partial _{a}\stackrel{*}{=}\delta ^{a}_{b}\partial _{a}$ is the unit vector field tangent to the geodesics orthogonal to $\Sigma $. (Physically this field represents the 4-velocity field of the set of observers which move along these geodesics.)

By using this unit timelike field one can easily introduce (in a neighbourhood of $\Sigma $) the Riemannian (Finsler-like) metric $h_{ab}=2v_{a}v_{b}-g_{ab}$ (with fixed $\vec{v}$):
\begin{align}
h_{ab}=(2v_{a}v_{b}-g_{ab})\stackrel{*}{=}2g_{a0}g_{b0}-g_{ab}.
\end{align}
From (30) it follows that the line elemnet determined by $h_{ab}$ reads
\begin{align}
\widetilde{ds}^{2}=h_{ab}dx^{a}dx^{b}\stackrel{*}{=}dt^{2}+\gamma _{\alpha \beta }dx^{\alpha }dx^{\beta }.
\end{align}
In NGC one can easily find the geometric structure imposed on a solution to the Einstein equations by the Finsler-like metric $h_{ab}$ and compare it with the original Lorentzian structure imposed by $g_{ab}$ e.g., for Riemannian tensor one has:
\begin{align}
&R_{\alpha \beta \gamma \delta }(g)=R_{\alpha \beta \gamma \delta }(\gamma )-\frac{1}{4}(g_{\alpha \delta ,0}g_{\beta \gamma ,0}-g_{\alpha \gamma ,0}g_{\beta \delta ,0}),
\\&R_{\alpha \beta 0 \delta }(g)=\frac{1}{2}(\mathcal{D}_{\beta }g_{\alpha \delta ,0}-\mathcal{D}_{\alpha }(\gamma )g_{\beta \delta ,0}),
\\&R_{\alpha 00 \beta }(g)=R_{0\alpha \beta 0 }(g)=\frac{1}{2}g_{\alpha \beta ,00}-\frac{1}{4}g^{\rho\sigma }g_{\alpha \rho ,0}g_{\beta \sigma ,0},
\\&R_{\alpha \beta \gamma \delta }(h)=R_{\alpha \beta \gamma \delta }(\gamma )+\frac{1}{4}(g_{\alpha \delta ,0}g_{\beta \gamma ,0}-g_{\alpha \gamma ,0}g_{\beta \delta ,0}),
\\&R_{\alpha \beta 0 \sigma }(h)=\frac{1}{2}(\mathcal{D}_{\beta }g_{\alpha \sigma ,0}-\mathcal{D}_{\alpha }(\gamma )g_{\beta \sigma ,0}),
\\&R_{\alpha 00 \beta }(h)=R_{0\alpha \beta 0 }(h)=\frac{1}{2}g_{\alpha \beta ,00}-\frac{1}{4}g^{\rho\sigma }g_{\alpha \rho ,0}g_{\beta \sigma ,0}.
\end{align}
$\mathcal{D}_{\alpha }(\gamma )$, $\mathcal {D}_{\beta }(\gamma )$ mean the covariant derivatives with respect to spatial metric $\gamma _{ab}\stackrel{*}{=}\gamma _{\alpha \beta }=-g_{\alpha \beta }$. (The covariant derivatives inside $\Sigma $).

\underline{Conclusion}

We have presented an universal method which attaches Riemannian metric and Riemannian structure to a given Lorentzian metric and structure. This method uses real quantities only.

So it seams better than the Wick rotation [4] which uses complex time. The method is also more general.

Solving Einstein equations  in terms of the metric $h_{ab}(x)$ one can find the spacetime metric $g_{ab}(x)$ from the equations $g_{ab}=2v_{a}v_{b}-h_{ab}$ [5].

\end{document}